\documentclass[runningheads]{cl2emult}

\usepackage{makeidx}  
\usepackage{graphicx} 
\usepackage{subeqnar} 
\usepackage{multicol} 
\usepackage{cropmark} 
\usepackage{math}     
\makeindex            

\begin{document}
\title*{Why Financial Markets Will Remain Marginally Inefficient}

%
%
%
%
%
\author{Yi-Cheng Zhang}
%
%
%
\institute{Institut de Physique Th\'eorique,
Universit\'e de Fribourg P\'erolles, Fribourg CH-1700, 
Switzerland}

\maketitle              

\begin{abstract}
I summarize the recent work on market (in)efficiency, highlighting key elements
why financial markets will never be made efficient. My approach is not by adding more empirical
evidence, but giving plausible reasons as to where inefficiency arises and why it's not rational to arbitrage it away.
\end{abstract}

\section{The Debate} 
The debate of whether a competitive market is efficient has been a hot topic in the last few decades. Academics, 
practitioners and general public are divided roughly into two camps: believers and non-believers of the so-called Efficient Market Hypothesis (EMH). Both camps have piled up mountains of evidence, yet no consensus can be said to have ever reached.

Efficient markets are a natural consequence of neoclassical economics, little wonder most believers of EMH come from mainstream economics departments. Paul Samuelson, the most influential living economist, claimed to have `mathemathically proven' that competitive markets
are efficient. Championing this camp now is the Chicago finance professor Eugene Fama. Non-believers, on the other hand, most are practitioners in the finance industry, and they make a living out of feeding on the residual inefficiency left in the market. Their very existence, they'd say, is proof enough of the contrary of EMH. One may get some idea from the recent book
{\it Alchemy of Finance} of George Soros. Academic economics is not a tight block, forceful 
opposition to EMH can be found in the writings of well-known economists such like Colin Camerer, Robert Shiller and especially Richard Thaler. Their approach is labelled sometimes as `behavorial economics'.

One has to look for roots in the neoclassical economics. Its founding doctrine is that economics is about optimally allocating scarce resources among competing uses. According to it market inefficiency, if any, would be arbitraged away by competitive players whose greed assures that 
markets stay at least closely to the perfect efficiency. It is later conceded that small inefficiency
can be still made consistent with EMH, since market players have to compensate their information
seeking and transaction costs. 
Representative work can be found in the recent book by Andrew Lo and Craig Mackinlay {\it A Non-Random Walk Down Wall Street}. Another concession from the mainsteam economics is that `smart money' enjoys some extra gains since there are `noise traders' foolish enough to sustain the fat cats on Wall Street.   

\section{Marginally Inefficient Markets (MEM) Theory}

Recently I introduced an approach taking a different perspective\cite{zha}. The so-called MEM theory maintains that financial markets are open systems, with unlimited potential entrants. Financial markets sport two main categories of players: producers and speculators . 

{\bf Producers} are the players who pay less attention than speculators to exploiting market inefficiency, their economic activities  outside the financial markets provide them better opportunities for profit than exploiting the 
market inefficiency. Yet their normal business conduct depends on heavily using the financial markets and their participation inadvertently injets the the elememts of predictability into the
markets. They do this not because of generosality but of inevitability. And they care less since
their core business is the outside economy.  

{\bf Speculators} are the players whose speciality is in `scavenging' financial markets for the tiny predictability (hence inefficiency) left by others. They do this purely for their own greed and inadvertently they render a social service: they provide liquidity so that the producers can use the market in a large scale under a relatively efficient conditions (compared to still larger opportunities in outside economy). Why don't our speculators arbitrage away the residual inefficiency in the financial markets? The key lies in that the predictability elements
in the financial markets are of probabilistic nature, no one can take out of the money without
making huge bets and taking on substanstial risks. In a previous work I show that to completely
arbitrage away the {\it probabilistic} predictability one needs infinite amount of risk capital. When the residual inefficiency is reduced to be small enough, the gain to risk ratio doesn't warrant further playing the arbitrage game. The possibility of arbitraging away inefficiency completely exists only in theoretical fiction.

{\bf Symbiosis} can be said of the relationship between producers and speculators. In the pursuit of their strictly selfish gains, due to their separate specialities (producers in outside economy and speculators in refining arbitrage), they unwillingly support each other's existence. A producer would, indeed sometimes does, avoid being taken advantage of by the speculators, by taking measures mimicking that of speculators. For instance it may avoid let traders to know its intentions of the forthcoming financial operations. But their ability of doing so is limited, not because competence, but rational thinking would let it concentrate on doing whatever it's good at and pay less attention to the relatively small market inefficiency. A speculator on the other hand, would like to make away money as much as possible,
but without risking substantial amount of money this is impossible. It has 
to be content with managing its risk while possibly make more money. There is a absolute limit in his ability in doing so.

Due to the impossibility of completely arbitraging away the inefficiency the two parties live in
a unwilling symbiosis bind. It's not necessarily a peaceful coexistence. For given the chance a speculator would like to scoop others' private information to grab a a big gain with little risk. But such easy gains are rare in competitive markets, especially when there is a credible institution enforcing fair-play rules. So they have to do the hard work of analyzing the tinest details
of available information and most importantly, comprehend the implications of myriads of confusing leads. There is a minimum service (liquidity) that a speculator must provide in order 
to extract a given amount of profit. Thus a properly designed institution can channel the greedy
speculators to do some social service by providing liqudity, while doing well.

On the other hand given the chance a producer would like to reduce their `loss' to the other players in the markets. But competence for it is different, and outside economic activities 
demand financial market participation at a rythm in general not synchrontized with that of 
financial markets. A producer must pay more
heed to the intrinsic production rythm, rather than to the market fluctuations. So the producers'  outside activities provide the market predictability in the first place, the speculators scavenging on these opportunities help provide liquidity.

It's possible that some of the speculators believe they are making money and in reality they actually lose. They can be said to be `noise traders'. In the market ecology it's likely that all types of players are intertwined. But I believe the fundamental element must lie in the symbiosis of the producers and speculators.

\section{Minority Game as a Test}

It's hard to test the above theory directly, as we cannot halt the normal financial operations
to see who played what role. However, powerful analytic tools are available to test the theory.
A recently developed model called Minority Game can be made convenient for our purpose\cite{cmz}. Minority
Game is a platform which is flexible enough to model quite a large class of multi-agents problem. In MG model players have a finite number of alternative strategies, the simplest case being two alternatives. Players have the choice of using the alternatives according to the market fluctuations. In most cases, more alternatives, more average gain. If a player for some reason has less alternatives than the others, one can show that in general this handicap leads
to relative loss.

From the standard MG model we can divide the players into two classes: producers and speculators. To model the real producers we impose on our MG producers the handicap of only
a single strategy, i.e. no other alternatives and their operation in the MG market is like an
automaton. Our MG speculators are normal MG players with two alternative strategies. They have
the capability, as in the standard MG model, to adjust their use of alternative strategies according market fluctuations. With the handicap, it's is an extreme case to model the producers' inflexibility facing market fluctuations. As discussed above, producers get profit from the outside economic activities and they don't mind their possible relatively small
loss due to market fluctuations. Left with only one strategy it is not hard for the speculators to detect that the market price has some regularity in it. But it's still highly non-trivial price signal since each producer's fixed strategy is randomly drawn, and in principle they are all different from each other. Moreover, the population is mixed with speculators and producers.

Players with no alternative or less alternatives tend to be rigid or less agile than others. A 
price signal containing some regular elements implies to have some predictability in it. Any predictability in the price history is by definition market inefficiency. From information
theoretic point of view we say that the conditional {\it entropy} of the price signal is not maximized. Or, there is some negative entropy left in the price signal that a savvy speculator can detect and act on it. The following calculation is provided by Damien Challet, now a postdoc
in Oxford.

\verb|\centering\noindent\mpicplace{width cm}{height cm}|.

\begin{figure}[h]

\centering
\includegraphics[width=0.6\textwidth]{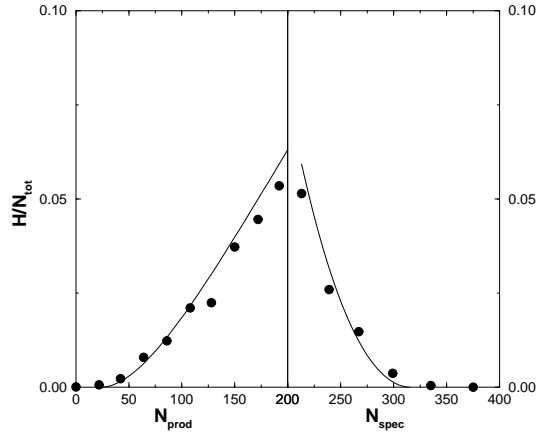}
\caption[]{The vertical axis is the negative entropy, any value above zero implying
some degree of predictability,
its special value 1 corresponds to certainty, which is normally
considered in economics literature. The original system consists of 200
pure speculators, on the horizontal axis we see with the gradual
introduction of producers, the negative entropy increases.
With their total number capped at 200, additional speculators are
introduced, they scavenge away the negative entropy.
Solid lines are analytical calculations, the dots are results from
directly simulating the model.}
\label{eps1}
\end{figure}

We consider a population initially only with pure speculators, they overwork for nothing since there is no negative entropy left in the system. In reality speculators would stop 
playing but in our model they are forced to play on. Now producers are gradually introduced and
the price signal gradually appears to be more and more interesting, in the figure we see the negative entropy rises signalling that the price becomes more and more predictable. We reach the
number of producers at 200 and keep them in the system. Now we want to introduce extra speculators into the system to take advantage of the information-rich price fluatuations. As we can see the predictability of the price, or the negative entropy starts to get reduced, until the
point when the market is made again efficient, i.e. null negative entropy. In our system the players are slaves and are forced to play as we pleased. In reality this mustn't be so since profit driven speculators would stop playing when the profitability becomes meager. Needless to say MG is but a toy model, and the
market impact is such that speculators can eliminate any inefficiency if their number is large enough. Contrary to real markets where infinite risk capital is needed.

\section{Conclusion}
It appears that previously studies on the market efficiency problem are concentrated on the markets themselves, insulated 
from the outside. Our novol approach emphasize that it's impossible to understand the problem in isolation. We need to 
cast the market mechanism into a bigger framework. In other words we are dealing an open system: there is always fresh
injection of predictable elememts from a fraction of players, who put in the `inefficiency' in the first place. This way 
of reasoning goes against the dominating neoclassical doctrine that systems can be reduced into mechanic pieces 
and understood separately. Technical tools that we have developped in recent years are quite advanced in dealing with 
systems of multi-agent and active work is still being performed. If markets are only marginally efficient, would we 
have a less elegant theory in place of EMH? Not a bit. It appears that the MEM theory outlined above is just part of a much
large theoretical framework, which will be discussed in full in a forthcoming book\cite{zha2}.
Acknowledgements: I benefitted from collaboration with Andrea Capocci, Damien Challet and Matteo Marsili.

\clearpage
\addcontentsline{toc}{section}{Index}
\flushbottom
\printindex

\end{document}